\documentclass[aps,prl,twocolumn,groupedaddress,epsfig]{revtex4}

\usepackage{epsfig}
\begin{document}



\title{No Froissart Bound from Gluon Saturation}

\author{Alexander Kovner$^a$ and
   Urs Achim Wiedemann$^b$}
\affiliation{$^a$Department of Mathematics and Statistics,
University of Plymouth,
2 Kirkby Place, Plymouth, PL4 8AA, U.K.\\
$^b$Theory Division, CERN, CH-1211 Geneva 23, Switzerland}

\date{\today}

\begin{abstract}
In the previous work~\cite{kw01}, we showed that while the
nonlinear QCD evolution equation of Balitsky and Kovchegov (BK)
leads to saturation of the scattering amplitude locally in impact
parameter space, it does not unitarize the total cross section.
This result was recently challenged by Ferreiro, Iancu, Itakura
and McLerran (FIIM) who claim that the dipole-hadron cross section
computed from the BK equation saturates the Froissart bound. In
this comment, we point to a fundamental error in the argument of
FIIM which invalidates their conclusion. We show that if the total
cross section violates unitarity for a coloured scattering probe,
it does so also for a colourless scattering probe.
\end{abstract}
\maketitle
 \vskip 0.3cm

In Ref.~\cite{kw01}, we demonstrated in two independent ways that
the dipole-hadron cross section evolved by the BK equation
\cite{bk} does not unitarize. The first derivation was given in
the target rest frame where the evolution resides in the
projectile wave function. This derivation uses explicitly {\it
colourless} scattering probes. The second derivation was given
within a formulation where the evolution is ascribed to the
target. This part of the argument can be construed as calculating
the scattering probability of a {\it coloured} probe. This latter
point seems to have caused some confusion. In particular, the
authors of Ref.~\cite{fiim02} claimed recently that our conclusion
holds for coloured probes only and that ``for the physically
interesting case where the external probe is a (colourless)
dipole, there is no problem with unitarity at all.'' The main
purpose of this note is to show that once the Froissart bound is
violated for a coloured probe, it is also violated for a
colourless scattering probe. We do this by considering the BK
evolution directly as the evolution of the target.

In the large $N_c$ limit, the BK equation is a closed equation
for the $t=\ln 1/x$ rapidity dependence of the scattering probability
$N(x,y)={1\over N_c}{\rm Tr}\langle 1-U^\dagger(x)U(y)\rangle $ of a
colour singlet dipole with charges at points $x$ and $y$,
\begin{eqnarray}
  &&{d\over dt}N(x,y) = \bar{\alpha}_s\,
    \int \frac{d^2z}{2\pi}\, {(x-y)^2\over(x-z)^2(y-z)^2}
    \label{kov} \\
  &&\qquad  [N(x,z)+N(y,z)-N(x,y)-N(x,z)N(z,y)] \, ,\nonumber
\end{eqnarray}
where $\bar{\alpha}_s = \frac{\alpha_sN_c}{\pi}$. The unitary
matrix $U(x)$ depends on the colour fields in the target and
represents the eikonal factor picked up by the fundamental charge
at transverse coordinate $x$ while propagating through the target
field.

In the following we concentrate on the evolution of the scattering
probability of a dipole of fixed size $r=x-y$ at fixed impact
parameter $b={x+y\over 2}$. We will sometimes write $N(r,b)$
instead of $N(x,y)$. Schematically the $x,y$ dependence of $N$ at
any given rapidity is depicted on Fig.1. The circle of radius
$R_0(t)$ bounds what one could call "the interior of the hadron".
For impact parameters $b\equiv {x+y\over 2}>R_0(t)$ the scattering
probability is vanishingly small for dipoles of any size. The
radius $R(t,Q^2)$ bounds the region inside which $N(r,b)\propto 1$
for fixed $r^2=Q^{-2}$ and outside which $N(r,b)\ll 1$. More precise
definitions of $R_0(t)$ and $R(t,Q^2)$ will be given shortly. The total
cross section for the scattering of a dipole of size $Q^{-1}$ on
the target is related to $R(t,Q^2)$ as
\begin{equation}
\sigma(t,Q^2)=2\int d^2b N(t,b,Q^{-1})\sim 2\pi R^2(t,Q^2)\, .
\label{crossec}
\end{equation}

\begin{figure}[h]
\epsfxsize=8.7cm
\centerline{\epsfbox{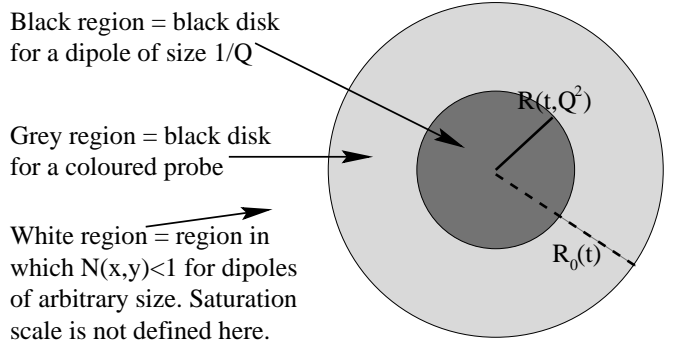}}
\caption{
The three impact parameter regions of a hadronic target, relevant
for the total dipole-hadron cross section
}\label{fig1}
\end{figure}

We follow the expansion of $R_0(t)$ and $R(t,Q^2)$ starting from some
initial value of rapidity $t_0$. In order to do this we place the initial
dipole at the impact parameter $b\gg R_0(t_0)$. We then follow the evolution
of $N(r,b)$.
We show that:

a). Initially, as long as $b\ge R_0(t)$ the main contribution to
the growth of $N(r,b)$ in eq.(\ref{kov}) comes from the points in
the central region $|z-x|=|z-y|\simeq b$ with $z<R_0(t)$ on the
right hand side of eq.(\ref{kov}). This long range contribution
leads to exponential growth
$R_0^2(t)\propto\exp\{\bar\alpha_s\epsilon t\}$. It also generates
a power like tail in the scattering probability $N(r,b)\propto
r^2/b^2$ and a nonvanishing saturation scale $Q_s(b)$ in the
faraway region of impact parameters.

b). After $R_0(t)$ reaches the impact parameter $b$, further
evolution in the vicinity of $b$ is dominated by local
contributions from $|z-y|\simeq |z-x|\ll b$. These do not affect
the growth of $R_0(t)$ but instead lead to a fast growth of
$Q_s(b)$. As a result it takes only a short time for $Q_s(t,b)$ to
reach the value $Q$, so that $R(t,Q^2)$ grows exponentially in
$t$.

We explain that the analysis of Ref.\cite{fiim02} fails on the
following point. The initial dipole in Ref.\cite{fiim02} is placed
in the "grey" region  at $R(t_0,Q^2)<b<R_0(t_0)$. This set up by
construction disregards the evolution in the regions of transverse
plane where the growth is dominated by the long range
contributions. It therefore {\it ab initio} misses the most
important contributions to the growth of the total cross section.
The crucial issue is not how fast the "grey" region becomes
"black", but instead how fast the "white" region becomes "grey".
By placing the initial dipole in the "grey" region, the authors of
Ref.\cite{fiim02} prevent themselves from addressing this
question.

We start by discussing some general properties of the scattering
probability $N(x,y)$.

\subsection{Properties of $N(x,y)$}
The dipole scattering probability $N(x,y)$ depends on the size $r
= x-y$ of the dipole $\lbrace x,y\rbrace$ and on its impact
parameter $b = \frac{1}{2}\left(x+y\right)$ with respect to the
hadronic target. For sufficiently small $b$ and $r$, it may be
characterized in terms of the saturation scale $Q_s(t,b)$
\begin{eqnarray}
  N(x,y) &=& {1\over N_c}{\rm Tr}\langle 1-U^\dagger(x)U(y)\rangle
  \nonumber\\
  &\simeq& 1 - \exp\left[ -Q_s^2(t,b)\, (x-y)^2\right]\, .
  \label{eq2}
\end{eqnarray}
It is important to keep in mind that the last equality only holds
for small impact parameters. This last equality implies that for
sufficiently large dipole size $r$, the scattering probability
always approaches unity. However in reality at large enough impact
parameters the scattering probability is perturbatively small for
dipoles of arbitrary size. This follows since the colour fields
far from the hadron vanish and thus $U(x) \simeq 1 + O(g)$. Hence
$N(x,y) \simeq O(\alpha_s) \ll 1$. Thus, for sufficiently large
impact parameter, there is no transverse scale $1/Q_s(t,b)$ above
which the dipole scattering probability is of order unity, in
contrast to the last line in (\ref{eq2}).

To clarify the limited applicability of the concept of saturation scale $Q_s(t,b)$
in characterizing the dipole scattering probability, we illustrate
the $r$- and $b$-dependence of $N(x,y)$ in more detail:

Consider first a small dipole $r \ll R_0(t)$ where $R_0(t)$
denotes the radius within which the target colour fields are large
at rapidity $t$. In this case, the two legs $x$ and $y$ of the
dipole see the target field of approximately the same normalized
strength $S(b)$, and the scattering probability can thus be
written in the factorized form
\begin{eqnarray}
  N(x,y) = n\left( r/L(b) \right)\, S(b)\, .
  \label{eq3}
\end{eqnarray}
Here, $L(b)$ characterizes the scale of correlations of the target
field (correlation length)
as a function of impact parameter. For impact parameter $b <
R_0(t)$ where target colour fields are large, the colour field correlation
length is equivalent to the inverse saturation scale, $L(b) = 1/Q_s(t,b)$.
For larger impact
parameters the correlation scale $L(b)$ is still a meaningful
concept but $Q_s(t,b)$ does not exist.

At initial rapidity $t_0$,
a physically sensible choice of the target field strength at
impact parameter $b$ is
\begin{eqnarray}
  S({b}) &=& 1\, \qquad \qquad \hbox{for}\quad  b < R_0(t_0)\, ,
  \label{eq4}\\
  S({b}) &\simeq& \exp\left[ - M\, (b-R_0(t_0)) \right]\,
  \quad \hbox{for}\quad  b > R_0(t_0)\, ,
  \label{eq5}
\end{eqnarray}
where $M$ is some mass scale. One should keep in mind, however,
that even though this {\it initial} profile shows an
exponential fall-off at large distances, there is no 
guarantee that this functional
shape is preserved in the evolution via (\ref{kov}). In fact, as
seen in eq. (\ref{eq8}) below, it is not.

As for the properties of the function $n$ in (\ref{eq3}), one
expects that it is a monotonically increasing function of its
argument. Since $N(x,y)$ is always smaller than unity and since
its maximal value is set by the strength of the target colour
field, we have $n(r/L(b)) \le 1$. For small dipoles, the scattering
probability is vanishingly small and grows  as $r^2$. Thus even
the central region of the target is transparent for sufficiently
small dipoles. For larger dipoles of size $r\propto L(b)$ the
central region is black. However, for the profile
(\ref{eq4},\ref{eq5}) the last line of (\ref{eq2}) holds only as
long as $S(b) = 1$, or $b\le R_0$. In short, the initial
scattering probability $N(t_0,x,y)$ can be parameterized  in terms
of some saturation scale $Q_s(t_0,b)$ only for impact parameters
lying in the ``grey'' region of the target, at $b< R_0(t_0)$. The
same of course applies also to scattering probability at any given
value of $t$.

The situation is very different for dipoles of large size where
the factorization (\ref{eq3}) does not hold. This is the case e.g.
for dipoles with $r = b$ sitting at large impact parameter $b \gg
R_0(t)$. If $\vec{r}$ and $\vec{b}$ are orthogonal, then both legs
of the dipole are in the ``white'' region (i.e., $x,y \gg R_0(t)$)
and the scattering probability is small, $N(x,y) \ll 1$. If
$\vec{r}$ and $\vec{b}$ are parallel instead, then one of the legs
is in the "grey" region (i.e., $x < R_0(t)$) while the other is in
the ``white'' region (i.e., $y \gg R_0(t)$, $U(y) \simeq 1$). As a
consequence, the dipole scattering probability loses in this
latter case any dependence on the size of the dipole and its
impact parameter. It can be viewed as the scattering probability
$1 - {1\over N_c}{\rm Tr}\langle U^\dagger(x)\rangle$ of a {\it
coloured} probe in the sense
\begin{eqnarray}
  N(x,y) &=& {1\over N_c}{\rm Tr}\langle 1-U^\dagger(x)U(y)\rangle
  \nonumber \\
  &\simeq& 1 - {1\over N_c}{\rm Tr}\langle U^\dagger(x)\rangle\, ,
  \quad \hbox{for}\quad y\gg R_0(t)\, .
  \label{eq6}
\end{eqnarray}
For $x<R_0(t)$ the target fields are large, and $<U(x)>=0$. Thus $N(x,y)=1$.
Accordingly, the radius $R_0(t)$ characterizes the
size of the black disk for a {\it coloured} probe. Again, for
dipoles for which (\ref{eq6}) applies, the scattering probability
$N(x,y)$ cannot be characterized in terms of a saturation scale
$Q_s$.

To sum up, the saturation scale $Q_s(t,b)$ is a meaningful concept
only in the region of impact parameter space which is black for
a coloured probe, $b < R_0(t)$.
For larger impact parameter, the
scattering probability is smaller than unity for {\it arbitrarily
large} dipoles and the last line of (\ref{eq2}) does not apply.

\subsection{Long-range Contributions violate the Froissart Bound}
We consider now a small-size dipole $r = x-y = 1/Q \ll R_0(t_0)$
which has at initial rapidity $t_0$ a negligible interaction
probability with the hadronic target since it sits at large impact
parameter $b \gg R_0(t_0)$. We ask how rapidly its scattering
probability grows to order unity as a function of rapidity $t$. We
first establish that the initial growth of $N(x,y)$ is dominated
by long-range contributions to the $z$-integral in (\ref{kov}).

If the newly generated dipole leg $z$ in eq.(\ref{kov})
sits at transverse position in the neighbourhood of $x$, $y$,
i.e. in the white region, then $N(x,z) \sim N(y,z) \sim N(x,y) \ll 1$
and the non-linear term in the BK equation can be neglected.
As a consequence, the short-range contribution grows proportional
to the local scattering probability which was taken to have an
exponential fall-off at initial rapidity,
\begin{eqnarray}
  \frac{d}{dt}N(x,y)\Bigg\vert_{\rm short}^{{\rm at}\,\, t=t_0}
  \propto
  \alpha_s\, e^{- M(b-R_0(t_0))}\, .
  \label{eq7}
\end{eqnarray}
For the long-range contribution, a lower bound of the growth of
$N(x,y)$ can be obtained by considering contributions from the
"grey" region only. With the initial dipole in the white region
(i.e. $N(x,y) \simeq 0$) but the leg $z$ in the "grey" region
(i.e. $N(x,z) \simeq 1$, $N(y,z) \simeq 1$), the BK equation
(\ref{kov}) gives
\begin{eqnarray}
  \frac{d}{dt}N(x,y)\Bigg\vert_{\rm long} =
  \bar{\alpha}_s\, \frac{r^2}{b^4}\, R^2_0(t)\, .
  \label{eq8}
\end{eqnarray}
Comparison of eqs. (\ref{eq7}) and (\ref{eq8}) shows that
at sufficiently large impact parameter $b \gg R_0(t)$, the
dominant growth of the dipole scattering probability comes from
long-range contributions.

It is important that  $R_0(t)$ appearing on the right hand side of (\ref{eq8})
is the radius of the black region for a {\it coloured} probe
in the precise sense specified in (\ref{eq6}). Namely, while the right hand
side of (\ref{kov}) depends on
colourless dipoles
$\lbrace x,y\rbrace$, $\lbrace x,z\rbrace$ and $\lbrace y,z\rbrace$,
only one of their {\it coloured} legs $z$ touches the target. Thus,
the scattering probability for these dipoles depends only on the
strength of the target-induced colour rotation
$1 - {1\over N_c}{\rm Tr}\langle U^\dagger(z)\rangle$ for this
one coloured leg, i.e. it is the scattering probability of
a {\it coloured} probe.

As shown in Ref.~\cite{kw01,kw02} and as rederived in
Ref.~\cite{fiim02}, the black disk radius $R_0(t)$ for a {\it
coloured} probe grows exponentially with rapidity,
\begin{equation}
  R_0(t)=R_0(t_0)\exp\hspace{-.1cm}\Big [\bar\alpha_s\epsilon (t-t_0)\Big ]\, .
  \label{eq11}
\end{equation}
Substituting this into eq.(\ref{eq8}) and integrating up to
rapidity
\begin{equation}
t_1={1\over\bar\alpha\epsilon}\ln{b\over R_0(t_0)}\label{t1}\, ,
\end{equation}
we find
\begin{eqnarray}
    N(r,b)\Bigg\vert_{t_1} \sim{r^2\over b^2}\, .
  \label{eq9}
\end{eqnarray}

Thus, as advertized we find that by the time the radius of the
"grey" region $R_0(t)$ reaches the impact parameter $b\gg
R_0(t_0)$, the scattering probability $N(r,b)$ is not
exponentially suppressed as suggested by eqs. (\ref{eq3},\ref{eq5})
but rather has a long power tail.

Since we expect $R(t,Q^2)<R_0(t)$, we can not use the
approximation eq.(\ref{eq8}) beyond the rapidity $t_1$. We can
nevertheless establish the exponential growth of $R(t,Q^2)$.
Starting at rapidity $t_1$, the point $b$ enters the "grey" region.
From this point on the evolution is dominated by "local"
contributions. The concept of saturation scale is well defined at
$b$ for $t>t_1$. As shown in many previous works \cite{levin}, and
rederived in \cite{fiim02} the saturation momentum grows
exponentially with rapidity
\begin{equation}Q_s(t,b) = Q_s(t_1,b)\, \exp\left[
\bar\alpha_s\, \lambda\, (t-t_1)\right]
\end{equation}
Thus at rapidity $t_2$ such that
\begin{equation}
t_2-t_1={1\over\bar\alpha\lambda}\ln{Q\over Q_s(t_1,b)}\label{t2}\, ,
\end{equation}
the scattering probability $N(r,b)$ reaches unity.

Putting eqs.(\ref{t1}) and (\ref{t2}) together we find that the
radius of the black disk $R(t,Q^2)$ reaches a given impact
parameter $b$ at rapidity $t$
\begin{equation}
t={1\over\bar\alpha\epsilon}\ln{R(t,Q^2)\over
R_0(t_0)}+{1\over\bar\alpha\lambda}\ln \left [{Q\over
Q_s(t_1,b)}\right ]\, .
\end{equation}
 We have not calculated the exact
value of $Q_s(t_1,b)$, but even without doing so we can put an
{\it upper} bound on $t$. Whatever the value of $Q_s(t_1,b)$, it
is greater than the inverse of the impact parameter
$b^{-1}$\cite{remark}. Thus substituting $b$ for $Q^{-1}(t_1,b)$
in eq.(\ref{t2}) and adding eq.(\ref{t1}) we find the upper bound
on the time that it takes the black disk $R(t,Q^2)$ to reach a
given impact parameter $b$
\begin{equation}
t<{1\over\bar\alpha\epsilon}\ln{R(t,Q^2)\over
R_0(t_0)}+{1\over\bar\alpha\lambda}\ln \left [R(t,Q^2)Q\right ]\, .
\end{equation}
Thus
\begin{equation}
\sigma(t,Q^2)\sim 2\pi R(t,Q^2)>2\pi
[R_0^\lambda(t_0)Q^{-\epsilon}]^{1\over
\epsilon+\lambda}e^{{\epsilon\lambda\over
\epsilon+\lambda}\bar\alpha_s t}\, . \label{rq}
\end{equation}
We stress that eq.(\ref{rq}) is the lower bound on $R(t,Q^2)$. For
example in the case when $Q_s(t_1)$ is a fixed $b$-independent
scale, the rates of growth of $R(t,Q^2)$ and $R_0(t)$ are the
same.

\subsection{Critique of FIIM}
The mistake in the argument of Ref. \cite{fiim02} can now be
easily understood. The set-up of Ref.~\cite{fiim02} is in
principle inadequate for studying the growth of $R(t,Q^2)$. In
Ref.~\cite{fiim02} the saturation scale at impact parameter $b$ is
assumed to be defined and to be perturbatively large at initial
rapidity $t_0$, $Q_s(t_0,b) \gg \Lambda_{\rm QCD}$. Thus, by
definition, this set-up ignores evolution in the region $b >
R_0(t_0)$ where the target field strength is initially small and
where the saturation scale does not exist at initial rapidity
$t_0$. As is clear from the discussion of eq. (\ref{eq8}) above,
the power tails generated in this region quickly start dominating
the total cross section, and therefore cannot be ignored. It is
this powerlike fall off of $N(b)$ (or equivalently $S(b)$) and not
the initial exponential one, that has to be taken as the "basic
distribution" which is further evolved by local effects.  This has
been ignored in Ref.\cite{fiim02}.

The physics is quite simple. We know from Refs.
\cite{kw01,kw02,fiim02} that $R_0(t)$ grows exponentially with
$t$. Then eq.(\ref{eq9}) tells us that for $R_0(t_0)<b<R_0(t)$,
that is up to distances exponentially larger than the initial
radius $R_0$, the long range contributions generate  a powerlike
tail for the scattering probability. This is to be contrasted with
the exponential fall off assumed for the initial profile. We can
now use the same argument as in Ref.\cite{fiim02} to establish
that the total cross section of a dipole of size $r=1/Q$ grows
exponentially. Starting from rapidity $t_1$, where $t_1$ is given
in eq.(\ref{t1}), the evolution of $N(r,b)$ at the impact
parameter $b$ is governed by the BFKL equation. The initial
condition for this BFKL evolution at $t_1$ is furnished by the
previous evolution from $t_0$ to $t_1$ which is dominated by the
long range contributions to eq.(\ref{kov}). This initial condition
is given by eq.(\ref{eq9}). Thus all we have to do is to replace
in eqs.(4.5) of Ref.\cite{fiim02} the exponential tail
$\exp[-2m_\pi b]$  by the power law tail ${r^2/b^2}=(Q^2b^2)^{-1}$
(see eq.(\ref{eq9})) and take into account the fact that the BFKL
evolution goes on not for the whole rapidity interval $t$, but
rather only for $t-t_1$. Neglecting the so called diffusion term
as in Ref.\cite{fiim02} we obtain:
\begin{equation}
N(Q^2,b)=\exp\left[-\ln [Q^2b^2]+\omega\bar\alpha_s
(t-t_1)-{1\over 2}\ln{Q^2\over\Lambda^2}\right]\, .
\end{equation}
Equating the scattering probability to one at $b=R(t,Q^2)$, we
obtain
\begin{equation}
R(t,Q^2)=\left[[{\Lambda\over
Q^3}]^{\epsilon}R_0(t_0)^\omega\right]^{1\over \omega+2\epsilon}
e^{{\omega\epsilon\over \omega+2\epsilon}\bar\alpha_s t}\, .
\label{rq1}
\end{equation}
Here $\Lambda$ is the saturation momentum in the central region of
the target at the initial value of rapidity. Given that
$2\lambda<\omega$, eq.(\ref{rq1}) is consistent with the lower
bound eq.(\ref{rq}). Thus we see again that the radius of the
black disk for the scattering of a colourless dipole grows
exponentially with $t$. The result of Refs.\cite{kw01,kw02} stands
confirmed.

\subsection{Final comments}

We finally give an intuitive physics argument of why the total
cross section of a colourless dipole violates the Froissart bound
if it is violated by a coloured scattering probe. To this end, we
first recall an argument due to Heisenberg: consider a theory with
a mass gap, where the profile of the distribution of matter
density in any target must decay exponentially at the periphery,
$\rho(b)\propto \exp\left[-mb\right]$. As this target is struck by
a projectile, in order to produce an inelastic scattering event at
least one particle must be produced. Assuming that the scattering
is local in the impact parameter plane, the region of the overlap
of the probe and the target must therefore contain energy at least
equal to the mass of the lightest particle, $m$. For scattering at
energy $E=s/m$ in the frame where all the energy resides in the
target, the target energy density is $E\rho(b)$. Thus the
scattering can only take place for impact parameters smaller than
those that satisfy $E\exp\left[-mb\right]=m$. Thus
$b_{max}={1\over m}\ln\left[s/m^2\right]$, which is equivalent to
the Froissart bound. Conversely, if the cross section grows as a
power of energy, then the density distribution in the target is
not exponential but power like. With $\rho(b)\propto b^{-\lambda}$
one obtains $b_{max}\propto s^{1\over\lambda}$ and this violates
the Froissart bound. Assume now that the total cross section for a
coloured probe (a quark, say) violates unitarity, i.e., the
scattering probability for a quark at impact parameter $b$ is
proportional to $\rho(b)\propto \frac{1}{b^{\lambda}}$. A
quark-antiquark dipole of transverse size $r$ will compare the
target field strength at two points separated by the distance $r$.
Its scattering probability is thus proportional to $r \frac{d}{db}
\rho(b) \propto r\, \frac{1}{b^{\lambda - 1}}$, which is still a
power-law fall-off. Hence, the total dipole-hadron cross section
violates the Froissart bound if the total quark-hadron cross
section does so.

\subsection{Note added}
After the appearance of the present paper, the authors of
\cite{fiim02} added a note to \cite{fiim02} agreeing with our
conclusions. Unfortunately they also attribute to the present
paper as well as to our previous papers \cite{kw01},\cite{kw02}
several statements which we never made nor intended to make.
For example, they attribute to our works "the statement
that one cannot compute the high-energy cross section using the
ideas of (perturbative) saturation". To the contrary, it is the 
main point of Ref.~\cite{kw02} to explain {\it how} the ideas of 
perturbative saturation can be used for the calculation of high-energy 
cross sections. For lack of space, we refrain here from giving
further examples of the same kind. Although Ref.~\cite{fiim02} is
quite misleading in the way it presents some aspects of our
work, we are confident that the careful reader will easily 
identify these inaccuracies.


\begin{thebibliography}{99}
\bibitem{remark} The value $Q^{-1}_s(t,b)\sim b$ is suggested by
eq.(\ref{eq9}).However since eq.(\ref{eq9})is the lower bound on
$N(r,b)$, the value $b$ is also a lower bound on $Q^{-1}(t,b)$.
\end{thebibliography}
\end{document}